\documentclass[aps, prd, twocolumn, lengthcheck, superscriptaddress, longbibliography, nofootinbib]{revtex4-1}

\usepackage{times}
\usepackage{amsmath,amssymb,amsfonts}
\usepackage{bm}
\usepackage{graphicx}
\DeclareGraphicsExtensions{.pdf,.eps,.png,.jpg}
\usepackage{acronym}
\usepackage{color}
\usepackage{xcolor}
\usepackage{tabularx}
\usepackage{multirow}
\usepackage{dcolumn}
\usepackage{soul}
\usepackage{float}
\usepackage{hyperref}
\usepackage{cleveref}
\usepackage{subfigure}
\usepackage{longtable}
\usepackage{tikz} 
\usetikzlibrary{shapes,arrows} 
\usepackage{soul}

\begin{document}

\title{Gravitational lensing aided luminosity distance estimation for compact binary coalescences}

\author{Kyungmin~\surname{Kim}$^{**}$}
\email{kkim@kasi.re.kr}
\affiliation{Korea Astronomy and Space Science Institute, 776 Daedeokdae-ro, Yuseong-gu, Daejeon 34055, Republic of Korea}

\author{Eungwang~\surname{Seo}$^{**}$}
\email{e.seo.1@research.gla.ac.uk}
\affiliation{SUPA, School of Physics and Astronomy, University of Glasgow, Glasgow G12 8QQ, United Kingdom}
\affiliation{Department of Physics, The Chinese University of Hong Kong, Shatin, New Territories, Hong Kong}

\author{Chunglee~\surname{Kim}}
\email{chunglee.kim@ewha.ac.kr}
\affiliation{Department of Physics, Ewha Womans University, 52 Ewhayeodae-gil, Seodaemun-gu, Seoul 03760, Republic of Korea}
\def\thefootnote{**}\footnotetext{K.K. and E.S. equally contributed to this work as co-first authors.}
\date[]{Last modified: \today}

\begin{abstract}
The luminosity distance is a key observable of gravitational-wave (GW) observations. We demonstrate how one can correctly retrieve the luminosity distance of compact binary coalescences (CBCs) if the GW signal is \emph{strongly lensed}. We perform a proof-of-concept parameter estimation for the luminosity distance supposing (i) strong lensing produces two lensed GW signals emitted from a CBC, (ii) the Advanced LIGO-Virgo network detects both lensed signals as independent events, and (iii) the two events are identified as strongly lensed signals originated from the same source. Taking into account the maximum magnification allowed in two lensing scenarios and simulated GW signals emitted from four different binary black holes, we find that the strong lensing can improve the precision of the distance estimation of a CBC by up to a factor of a few compared to that can be expected without lensing.
\end{abstract}


\maketitle

\section{Introduction}

The luminosity distance $D_L$ of a gravitational-wave (GW) source is one of the direct observables available with GW observations. It has significant implications in astronomy as well as astrophysics. Distance estimation to a GW source that can be bright in electromagnetic waves is invaluable for follow-up observations to design and analyze the follow-up observations. Distance information along with the sky locations of GW sources is also important to understand the formation, evolution, and underlying properties of the source (e.g., see ~\citep{KAGRA:2013rdx} and references therein). 

In the context of GW observation with the network of the Advanced LIGO (aLIGO)~\cite{TheLIGOScientific:2014jea} and the Advanced Virgo (AdV)~\cite{TheVirgo:2014hva}, the precision of the distance estimation---in terms of the width of the posterior distribution of estimated distance---of a compact binary coalescence (CBC) is, in general, subject to the signal-to-noise ratio (SNR) of the signal given detector sensitivity and noises contained in  data~\cite{Christensen:2022bxb}. In practice, SNR is proportional to the strain amplitude of a GW signal and inversely proportional to the square root of the detector noise power spectral density (PSD). If a GW signal is detected by template-based search methods, e.g., \textsc{PyCBC}~\cite{DalCanton:2014hxh,Usman:2015kfa,alex_nitz_2020_4075326}, \textsc{gstlal}~\cite{Messick:2016aqy,Sachdev:2019vvd}, \textsc{SPIIR}~\cite{spiir:2012prd,Chu:2020pjv}, \textsc{MBTA}~\cite{MBTA:2016cqg}, and \textsc{IAS}~\cite{Venumadhav:2019tad}, with a higher SNR by some signal enhancement mechanisms, it is expected that the precision of the parameter estimation (PE) for the CBC can be improved. 

Under a given detector sensitivity, a possible astrophysical phenomenon of obtaining higher SNR for a GW signal is that the GW signal experiences strong lensing~\cite{Ohanian:1974, Bliokh:1975apss, Bontz:1981apss, Thorne:1983, Schneider:1992, takahashi:2003apj}. It is anticipated in the literature that strong lensing of GW results in not only magnifying an original GW signal but also producing multiple GW ``images'' (lensed GW signals, hereafter) having different magnifications.
This means that an \emph{apparent} luminosity distance estimated from each lensed signal can be different from a \emph{true} distance---the distance to be measured when there is no lens system between the observer and the source---to the source. 

Like the GW signals hitherto detected, multiple-lensed GW signals are to be observed as transient events with different SNRs at different observation times~\cite{wang1996gravitational,dai2017waveforms,ng2018precise,li2018gravitational,oguri2018effect}. If all multiple-lensed GW signals from a CBC are detected and their physical association is identified, combining lensed signals in the PE analysis and estimating true parameters of the source are available~\cite{janquart2021fast,lo2023bayesian,janquart2023return}. For example, \cite{janquart2021fast} discussed that utilizing multiple-lensed GW signals is helpful in the sky localization of host galaxies of original unlensed GW signals. Furthermore, the enhanced signals expected by strong lensing make it possible to reduce uncertainties in the parameter estimation (PE).
Therefore, strong lensing of GWs can be a promising scenario not only for enhancing the SNR of observed GW signals but also for inferring the same CBC parameters, such as $D_L$, with reduced uncertainties.

Motivated by the discussed advantages of strong lensing in the estimation of distance to CBCs and forecast studies~\cite{ng2018precise,li2018gravitational,oguri2018effect,mukherjee2020probing,wierda2021beyond} on observing strongly lensed GWs in the near future, we explore the feasibility of utilizing two lensed GW signals for more precise estimation of the true $D_L$ to a CBC. We demonstrate the PE analysis supposing an equal-mass binary black hole (BBH). We examine the best possible precision in estimating the true $D_L$ to the BBH, taking into account the advanced GW detector network sensitivity. Posterior distributions of $D_L$ are obtained from the PE strategy described in this paper. We also test the feasibility with three other mock BBHs that are similar to the selected BBHs---representing different BBH populations and having different SNRs for their signals---listed in the latest gravitational-wave transient catalog, GWTC-3~\cite{LIGOScientific:2021djp}.
For all simulated signals, we consider two examples of detector sensitivities with different assumptions on the noise realization.

In this work, we obtain a posterior probability density function (PDF) of true $D_L$, $p(D_L)$, assuming the GW signals from the BBHs are lensed or unlensed, respectively. 
Looking at the $p(D_L)$ and the width of 99\% credible interval, it turns out that the method presented in this paper enables us to successfully retrieve the true $D_L$ from different apparent luminosity distances to strongly lensed GW signals.
We conclude that if multiple lensed GW signals, i.e., strong lensing counterparts of an original signal, are detected, the true $D_L$ to a GW source can be better constrained compared to what can be expected without any gravitational lensing effects on the original signal.

We organize this paper as follows: In Sec.~\ref{sec:lensing_basics}, we briefly summarize analytic formulations for the strong lensing of GWs adopted in this work. In Sec.~\ref{sec:methods}, we describe the procedure of parameter estimation for strongly lensed GWs. In Sec.~\ref{sec:results}, we present the posterior probability density of $D_L$ associated with and without lensing scenarios under different detector sensitivities and/or noise realizations. Finally, we discuss the result of this work and its possible implications in Sec.~\ref{sec:discussions}.

\section{Strong lensing of GW signals}
\label{sec:lensing_basics}

We adopt the lens configuration described in \cite{takahashi:2003apj}. We consider a lens located between a BBH and an observer, i.e., the ground-based GW detector network. We assume two strongly lensed GW signals are generated and propagated toward the detector network as an originally unlensed GW signal radiated from a BBH passes through the lens. In this work, we suppose a galaxy-like lens and apply the thin-lens approximation. Then we can obtain the lensed GW signal $h_l(f)$ from $h_u(f)$ by a simple relation: 
\begin{equation}
h_{l}(f)\!=\!F(f) h_{u}(f)~, \label{eq:lensedGW}
\end{equation}
where $F(f)$ is an amplification factor that determines the lensing characteristics. As we consider a galaxy-like lens, $F(f)$ can be obtained in the geometrical optics limit. 

In this work, we consider two lens models, the point-mass (PM) and the singular isothermal sphere (SIS). An amplification factor for both lens models is given as 
\begin{equation}
F(f)\!=\!\sqrt{|\mu_{+}|}\!-\!i \sqrt{|\mu_{-}|} e^{2\pi i f \Delta t}~. \label{eq:amplification}
\end{equation}
Here, $\mu_{+}$ and $\mu_{-}$ are individual magnification factors corresponding to each lensed GW signals $h^\textrm{I}_l(f)$ and $h^\textrm{II}_l(f)$, respectively. 
Also, $\Delta t$ is the time delay between the arrival times of $h^\textrm{I}_l(f)$ and $h^\textrm{II}_l(f)$ to an observer. For each lens model, $\mu_\pm$ and $\Delta t$ can be written as follows:
\begin{eqnarray}
\textrm{PM:}~&\mu_\pm~&= \frac{1}{2} \pm \frac{y^2 + 2}{2y\sqrt{y^2 + 4}}~, \nonumber\\
\textrm{SIS:}~&\mu_\pm~&= \pm1 + \frac{1}{y} \quad \text{for} \; y < 1~,\label{eq:mu_pm}
\end{eqnarray}
and
\begin{eqnarray} 
\textrm{PM:}~&\Delta t~&= \frac{4 G M_{lz}}{c^3} \left[ \frac{y \sqrt{y^2\!+\!4}}{2} + \ln \left\{ \frac{\sqrt{y^2\!+\!4}+y}{\sqrt{y^2\!+\!4}-y} \right\} \right]~, \nonumber\\
\textrm{SIS:}~&\Delta t~&= \frac{8 G M_{lz} y}{c^3}~. \label{eq:dt}
\end{eqnarray}
In Eqs.~\eqref{eq:mu_pm} and \eqref{eq:dt}, $y$ denotes the parameterized source position following the lens configuration used in \cite{takahashi:2003apj}. The range of $y$ can be constrained to be $[0.1,~1.0)$ in this work. The expected occurrence rate of strongly lensed GWs sets the lower limit of $y\!\geq\!0.1$~\cite{Lai:2018prd,Hannuksela:2019kle,LIGOScientific:2021izm,Liu_2021,LIGOScientific:2023bwz}. The upper limit is given by the SIS model, i.e., the $y$-dependent validity of $F(f)$, requires $y\!<\!1$ in order to produce two lensed signals with SIS~\cite{takahashi:2003apj}. The expression of $F(f)$ implies that PM always produces two lensed signals with any $y$. As shown in Eq.\ \eqref{eq:dt}, the time delay is proportional to a redshifted mass $M_{lz}\!=\!M_l (1+z_l)$ of a lens at redshift $z_l$. As a representative value, we set $M_{lz}\!=\!10^{11.5}M_\odot$, which results in a time delay from weeks to months between $h^\textrm{I}_{l}(f)$ and $h^\textrm{II}_{l}(f)$ in the range of $y$ considered in this work.

\section{Methods} \label{sec:methods}
\subsection{Assumptions used in PE}
There are a few assumptions we made in this work about the source and the lensing phenomenon: (a) A BBH originally radiates $h_u(f)$, (b) a lens model and a GW waveform model for the source are known, (c) the lensed signals $h^\textrm{I}_{l}(f)$ and $h^\textrm{II}_{l}(f)$ are characterized by $\mu_{+}$ and $\mu_{-}$, respectively, and (d) both $h^\textrm{I}_{l}(f)$ and $h^\textrm{II}_{l}(f)$ are detected and their physical associations are identified. 

The IMRPhenomXPHM waveform model~\cite{Pratten:2020ceb} is used for the preparation of both the injection signal and template signal. This model is phenomenological and can describe the full inspiral-merger-ringdown phases of the GW signal from a CBC. It also allows us to simulate GW signals from a precessing CBC. By using the same waveform model for the injection and template, we can ignore systematic biases that can possibly arise from the inconsistency between the template and injected signal.

\begin{table}[t!]
\caption{Selected parameters and prior assumptions used to generate an unlensed GW signal $h_u(f)$ from an equal-mass BBH.
\label{tab:inj_params}}
    \centering
    \begin{tabularx}{1\linewidth}{@{} l @{\extracolsep{\fill}} *{3}{c} @{}}
        \toprule
        \multirow{2}{*}{Parameter} &
        \multirow{2}{*}{Unit} & \multirow{2}{*}{Value} & Prior \\
        & & & distribution \\
        \hline
        Component masses, $m_1$ \& $m_2$ & $M_\odot$ & 30.0 & Uniform \\
        Chirp mass, $\mathcal{M}$ & $M_\odot$ & 26.1 & Uniform \\
        Luminosity distance, $D_L$ & Gpc & 3 & Euclidean \\
        Right Ascension & rad & 1.3750 & Uniform\\
        Declination & rad & -1.2108 & Isotropic\\
        \hline
        \hline
    \end{tabularx}
\end{table}

In addition, we assume the design PSDs~\cite{abbott2020prospects} of the detector network consisting of the aLIGO-Hanford~(H), the aLIGO-Livingston~(L), and the AdV~(V) (the HLV detector network hereafter). For example, the network SNR of the $h_u(f)$ of the equal-mass BBH is obtained to be 12 based on this assumption.

Assumptions of priors of parameters used in the PE analysis are as follows. A uniform distribution in an Euclidean volume~\cite{Veitch:2014wba} is used for the distance prior with a range of $[0.1,10]~\textrm{Gpc}$.
For other source parameters, including chirp mass $\mathcal{M}$ with a range of $[10,50]~M_{\odot}$, we adopt the default precessing BBH prior distributions implemented in the \textsc{Bilby} library~\cite{Ashton:2018jfp, romero2020bayesian}. Some of the source parameters and prior assumptions used for the equal-mass BBH considered in this work are summarized in Table~\ref{tab:inj_params}.

\subsection{PE for unlensed signals}
Let us consider an unlensed signal $h_u(f)$ as the true GW signal from a CBC observable when there is no lens between the source and an observer. Utilizing the \textsc{Bilby} library and \textsc{Dynesty} nested sampler~\cite{dynesty_article,nested_sampling}, we perform PE (labeled as `normal PE' in Fig.~\ref{fig:flowchart}) for a simulated $h_u(f)$. 

Fig.~\ref{fig:unlensed} presents $p(\mathcal{M})$ and $p(D_L)$ for $h_u(f)$ obtained from the PE analysis of the $h_u(f)$ from the equal-mass BBH. In order to yield the best possible precision for the distance estimation via GW observation of a given detector sensitivity, we assume a ``zero-noise'' realization~\cite{Rodriguez:2013oaa, Favata:2021vhw}. The injected values of $\mathcal{M}$ and $D_L$ are successfully retrieved within the 99\% credible interval (C.I.) as expected. The $p(D_L)$ presented in Fig.~\ref{fig:unlensed} is then used as the reference to be compared to $p(D_L)$ from $h_l^\textrm{I}(f)$ and $h_l^\textrm{II}(f)$ following the flowchart shown in Fig.\ \ref{fig:flowchart}.

\subsection{PE for lensed signals}

\tikzstyle{decision} = [diamond, draw, fill=blue!20, text width=4.0em, text badly centered, node distance=2cm, inner sep=0pt]
\tikzstyle{block} = [rectangle, draw, fill=blue!20, text width=5em, text centered, rounded corners, minimum height=3em,text width=5.5em,node distance=2cm]
\tikzstyle{arrow} = [draw, -latex',thick]
\tikzstyle{cloud} = [draw, ellipse, fill=red!20, node distance=3cm, minimum height=2.5em]
\tikzstyle{circle} = [draw, ellipse, fill=green!20, node distance=1cm, minimum height=2em,text width=3em,text centered]
\begin{figure}[t]
    \centering
\begin{tikzpicture}[node distance = 0cm, auto]
\node [circle] (init) {galactic lens};
\node [cloud, left of=init] (Start) {$h_{u}(f)$};
\node [cloud, right of=init] (Start2) {$h^\mathrm{I}_{l}(f)$,~$h^\mathrm{II}_{l}(f)$};
\node [block,below of=init] (init2) {Retrieval \\process};
\node [block, below of=Start2] (init4) {Joint PE};
\node [block, below of=Start] (init3) {Normal PE};
\node [decision, below of=init4] (End3) {$D^{\rm PM,SIS}_{L,\pm}$};
\node [decision, below of=init2] (End) {$D^{\rm PM,SIS}_{L}$};
\node [decision, below of=init3] (End2) {$D^{\rm UL}_{L}$};
\path [arrow] (Start) -- (init);
\path [arrow, dashed] (init3) -- (End2);
\path [arrow, dashed] (init4) -- (End3);
\path [arrow, dashed] (Start) -- (init3);
\path [arrow, double] (init) -- (Start2);
\path [arrow, dashed] (Start2) -- (init4);
\path [arrow, dashed] (End3) -- (init2);
\path [arrow, dashed] (init2) -- (End);
\path [arrow, double, dotted] (End) -- node [text width=2cm,above=0.7em ] {~~comparison} (End2);
\path [arrow, double, dotted] (End2) -- (End);
\end{tikzpicture}
\caption{Flowchart of parameter estimation for a strongly lensed GW signal. The posterior of $D^{\rm UL}_{L}$ (distance to an unlensed signal $h_{\textrm{u}}$($f$)) is calculated first. Then,  posteriors of $D^{\rm PM, SIS}_{L}$, distances to the two lensed signals from a given CBC, are computed from the joint PE and retrieval process as described in the text. Finally, the two distance posteriors are compared to obtain the posterior of the true distance $D_L$. This approach can be applied to any CBCs as long as two lensed signals are identified.}
\label{fig:flowchart}
\end{figure}
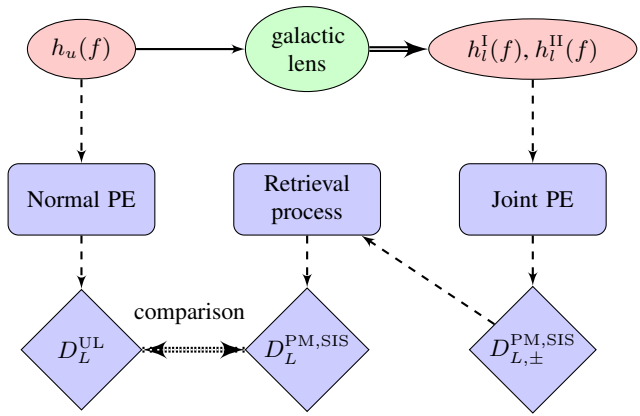

As a condition for the most optimistic strong lensing scenario, we focus on the maximum amplification available by the assumed lens models. For our lens models and the considered range of $y$, the maximum amplification occurs if $y\!=\!0.1$ by Eqs.~\eqref{eq:amplification} and \eqref{eq:mu_pm}. Hence, we focus on two lensed signals $h^\textrm{I,II}_l(f)$ with the maximum amplifications assuming $y\!=\!0.1$.

We conduct the PE analyses for the injected $h^\textrm{I,II}_l(f)$ following the procedure depicted in Fig.~\ref{fig:flowchart}: Utilizing the two lensed GW signals $h_l^\textrm{I,II}(f)$, we calculate a joint likelihood based on both lensed signals and conduct \emph{delensing} in order to retrieve the posterior PDF for the true luminosity distance $p(D_L)$ to a CBC. Following similar processes, other source parameters such as $\mathcal{M}$ can be also inferred. However, we only discuss the distance, the parameter of interest of this work.

\begin{figure}[t!]
\centering
\includegraphics[width=8.6cm]{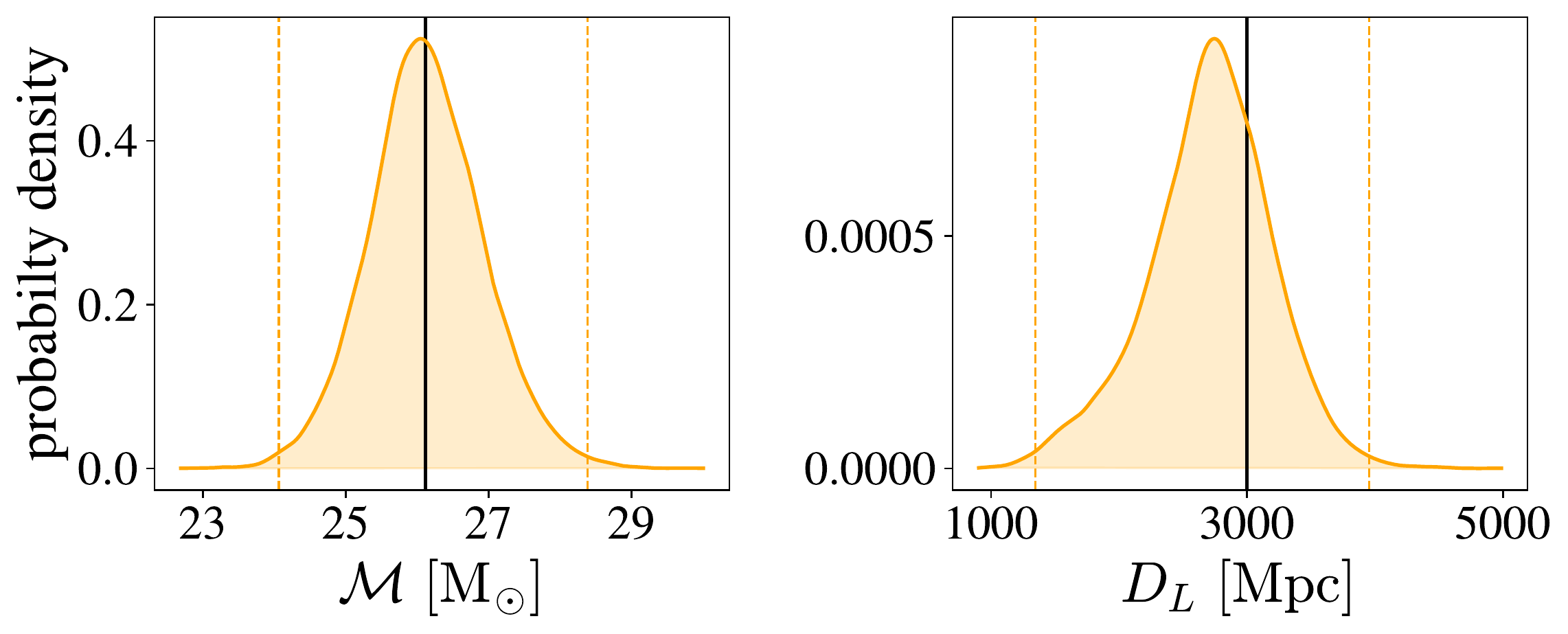}
\caption{Posterior PDFs of chirp mass $\mathcal{M}$ (the left panel) and luminosity distance $D_{L}$ (the right panel) recovered from the equal-mass BBH without lensing. Black solid lines represent injected parameters. Vertical orange dotted lines are the lower and upper bounds of the 99\% C.I., respectively. 
\label{fig:unlensed}}
\end{figure}

\begin{figure}[t!]
\centering
\includegraphics[width=8.6cm]{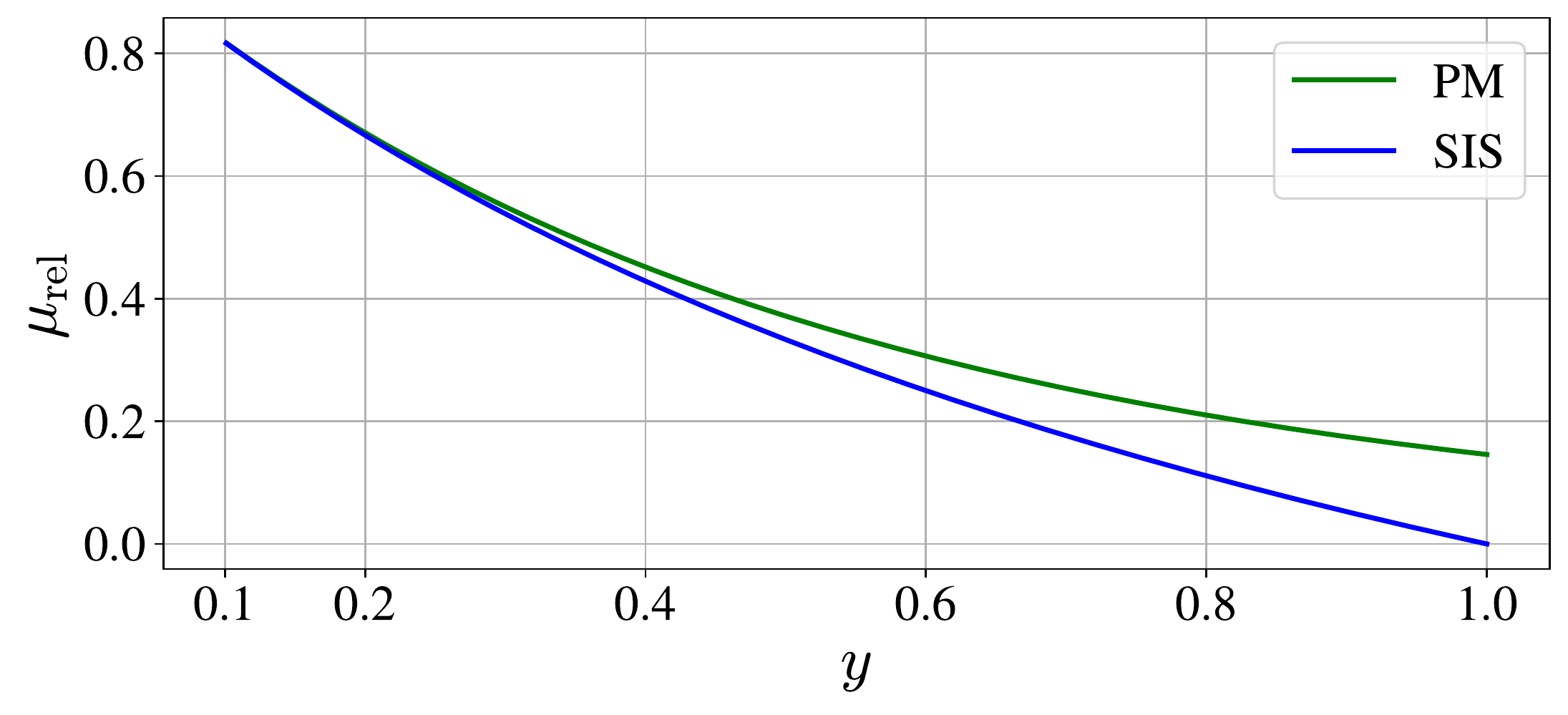}
\caption{Relation between the relative magnification factor $\mu_{\rm rel}$ and the source position $y$. The green and blue lines are obtained from the point-mass (PM) and singular isothermal sphere (SIS) models, respectively. 
\label{fig:relation_murel_y}}
\end{figure}

The \textsc{Golum} pipeline~\cite{janquart2021fast}, developed based on the \textsc{Bilby} library, enables us to infer the apparent parameters, e.g., $D_{L\pm}$ and $\mathcal{M}_{\pm}$ for $h^\textrm{I,II}_{l}(f)$, respectively. 
In addition, it allows us to infer lensing parameters such as the relative magnification factors $\mu_\textrm{rel}$, time delays $\Delta t$ between arrival times of lensed GW signals, and differences between Morse number $\Delta n$ of the lensed signals~\cite{dai2017waveforms}.
In particular, $\mu_\textrm{rel}$ can be rewritten by the apparent distances for both PM and SIS models in the geometrical optics limit, that is, $\mu_\textrm{rel}\!\equiv\!\left|\mu_{-}/\mu_{+}\right|\!=\!(D_{L+}/D_{L-})^{2}$. Fig.~\ref{fig:relation_murel_y} shows $\mu_\textrm{rel}(y)$ for PM and SIS.

Although we assume the physical association of $h^\textrm{I,II}_l(f)$ is identified, the two lensed signals $h^\textrm{I,II}_l(f)$ would likely be identified as two independent events separated by time in practical GW observation. 
Each lensed signal is then analyzed individually. If the SNRs of the two lensed signals are large enough, the most likely values of $\mathcal{M}$ obtained from two lensed signals would be almost identical within the uncertainty attributed to the sensitivity of the detector and analysis pipelines.
However, distance estimates from the two lensed signals are expected to be more different depending on individual magnification factors for each lensed signal. The relation between the \emph{apparent} luminosity distances and the true luminosity distance can be written as $D_{L\pm}\!=\!D_L / \sqrt{|\mu_\pm|}$.
In order to reflect this realistic observation scenario, we inject apparent distances $D_{L\pm}$ to simulate two lensed signals $h^\textrm{I,II}_l(f)$, respectively.

PE analyses for lensed signals involve assumptions on lensing parameters and $D_{L_{\rm \pm}}$, in addition to those used for an unlensed signal. Based on what is discussed earlier, we use the same injection parameters for 
  $h^\textrm{I,II}_l(f)$ with those used for $h_u(f)$ except $D_{L_{\rm \pm}}$. For example, the chirp masses of the equal-mass BBH for lensed or unlensed GW signals are assumed to be the same ($\mathcal{M}_+\!=\!\mathcal{M}_-\!=\!\mathcal{M}\!=\!26.1 M_\odot$). The apparent sky locations of two lensed signals are assumed to be the same because the subtle differences between the lensed signals in the sky cannot be distinguished by the sensitivities of the current advanced detector network (e.g., \citep{LIGOScientific:2021djp}). As for prior distributions for lensed signals, we use the same assumptions used for $h_u(f)$ and assume a uniform prior distribution for $\mu_\mathrm{rel}$~\cite{janquart2021fast}.
Note that we only consider $\mu_{\rm rel}$ among three lensing parameters ($\mu_{\rm rel}, \Delta t$, and $\Delta n$) for a given scenario, as the other two do not affect luminosity distance estimation.

We obtain the maximum likelihood value $\mu_\mathrm{rel,~max}$ from the $p$($\mu_\mathrm{rel}$) and find corresponding $y$($\mu_\mathrm{rel,~max}$) from Fig.~\ref{fig:relation_murel_y}.
Also, it is straightforward to calculate $\mu_\pm$ using Eq.~\eqref{eq:mu_pm}. When $\mu_\pm$ and $D_{L\pm}$ are at hands, we can obtain $D_L$ by $D_L\!=\!\sqrt{|\mu_+|}D_{L+}$ or $D_L\!=\!\sqrt{|\mu_-|}D_{L-}$. 

It is shown in~\cite{janquart2021fast} that we can better constrain the lensing and source parameters by \emph{reweighting} the posterior samples of individual lensed signals.
The \emph{reweighted} posterior PDFs can be obtained by combining 
posteriors of lensing parameters and apparent source parameters obtained in earlier steps (see Equation~(15) in~\cite{janquart2021fast} for an example of the reweighted posterior).
Then, $p(D_L)$ can be obtained from a reweighted posterior of the apparent distance $p(D_{L+})$ by $p(D_L)\!=\!\sqrt{\left|\mu_+\right|}p(D_{L+})$ (labeled as a retrieval process in Fig.~\ref{fig:flowchart}). The individual magnification factor $\mu_+(y)$ is determined by $y\!=\!y(\mu_{\rm rel,~max})$. 

In this work, we choose $h^\textrm{I}_l$ assuming (i) $h^\mathrm{I}_l$ arrives earlier than $h^\mathrm{II}_l$ and (ii) $h^\mathrm{I}_l$ is experienced stronger magnification than $h^\mathrm{II}_l$. We perform the same reweighting procedure---described in Equation (15) and Appendix A of~\cite{janquart2021fast}---and calculate posteriors of lensing and apparent source parameters of $h^\textrm{I}_l$. One can choose either $h^\textrm{I}_l$ or $h^\textrm{II}_l$ when combining the two likelihoods in order to determine the true distance posterior of the source. 

In order to examine the effect of noise as well as the capability of different observing runs, we consider two configurations that determine the network detector sensitivities: (i) the HLV design PSDs with zero-noise (Case A) and (ii) the HLV O3a PSDs~\footnote{The corresponding PSD data for H, L, and V can be found from https://dcc.ligo.org/LIGO-T2000012/public. Note that this PSD data is based on the first three months of the third observing run (O3).} with Gaussian noise (Case B).

\begin{figure}[t]
\centering
\includegraphics[width=\linewidth]{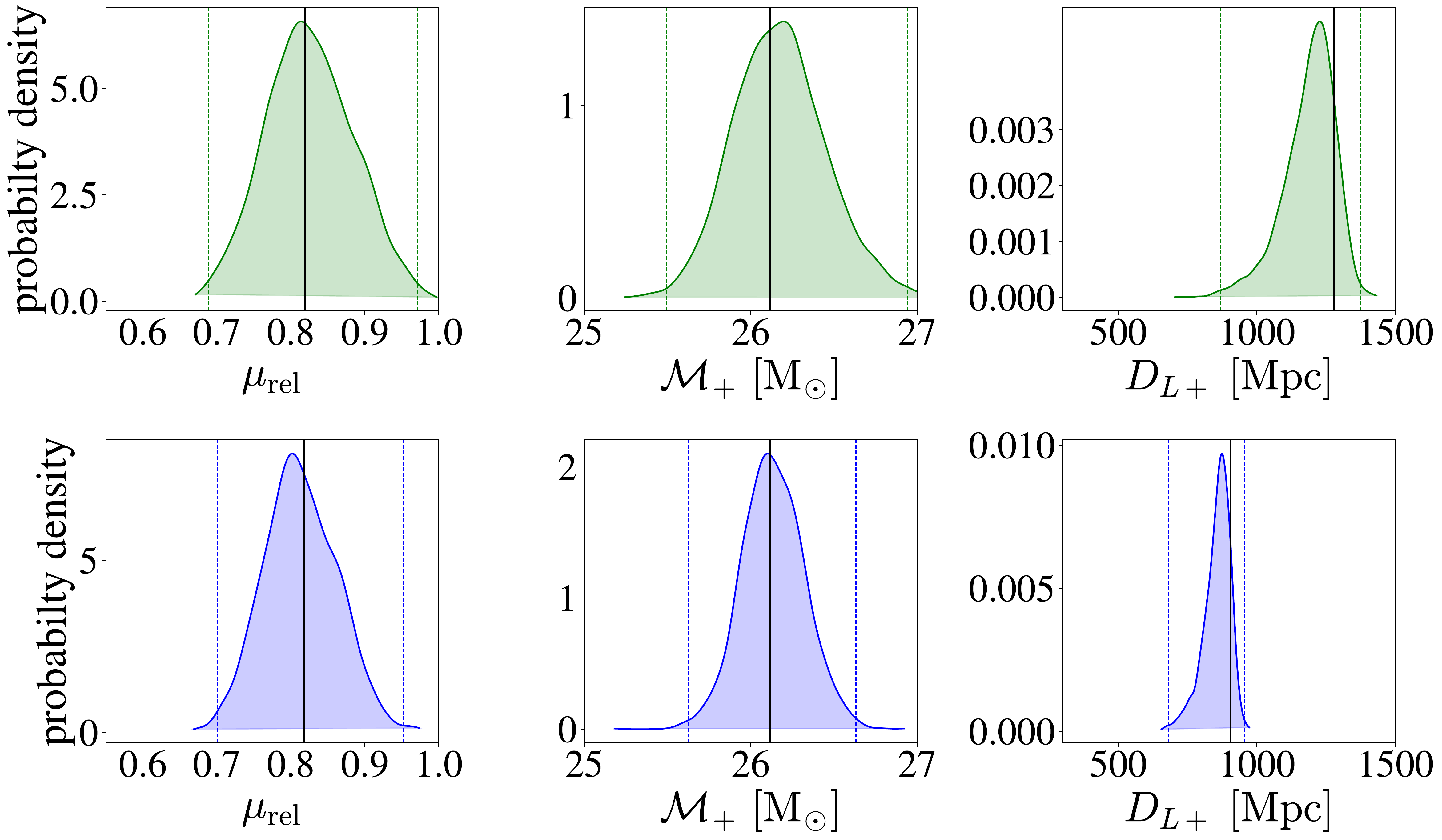}
\caption{Posteriors of $\mu_\mathrm{rel}$, $\mathcal{M}_{+}$, and $D_{L+}$ obtained for $h^\mathrm{I}_l$ based on PM (green, top) and SIS (blue, bottom) models obtained from the equal-mass BBH. Black solid lines indicate $\mu_\mathrm{rel}(y\!=\!0.1)$ and injected values of $\mathcal{M}_{+}$, and $D_{L+}$. Vertical dashed lines are the lower and upper bounds of the 99\% C.I.}
\label{fig:relative}
\end{figure}

\begin{figure*}[t]
\centering
   \subfigure[\; HLV design PSDs with zero noise]
{\includegraphics[width=.48\linewidth]{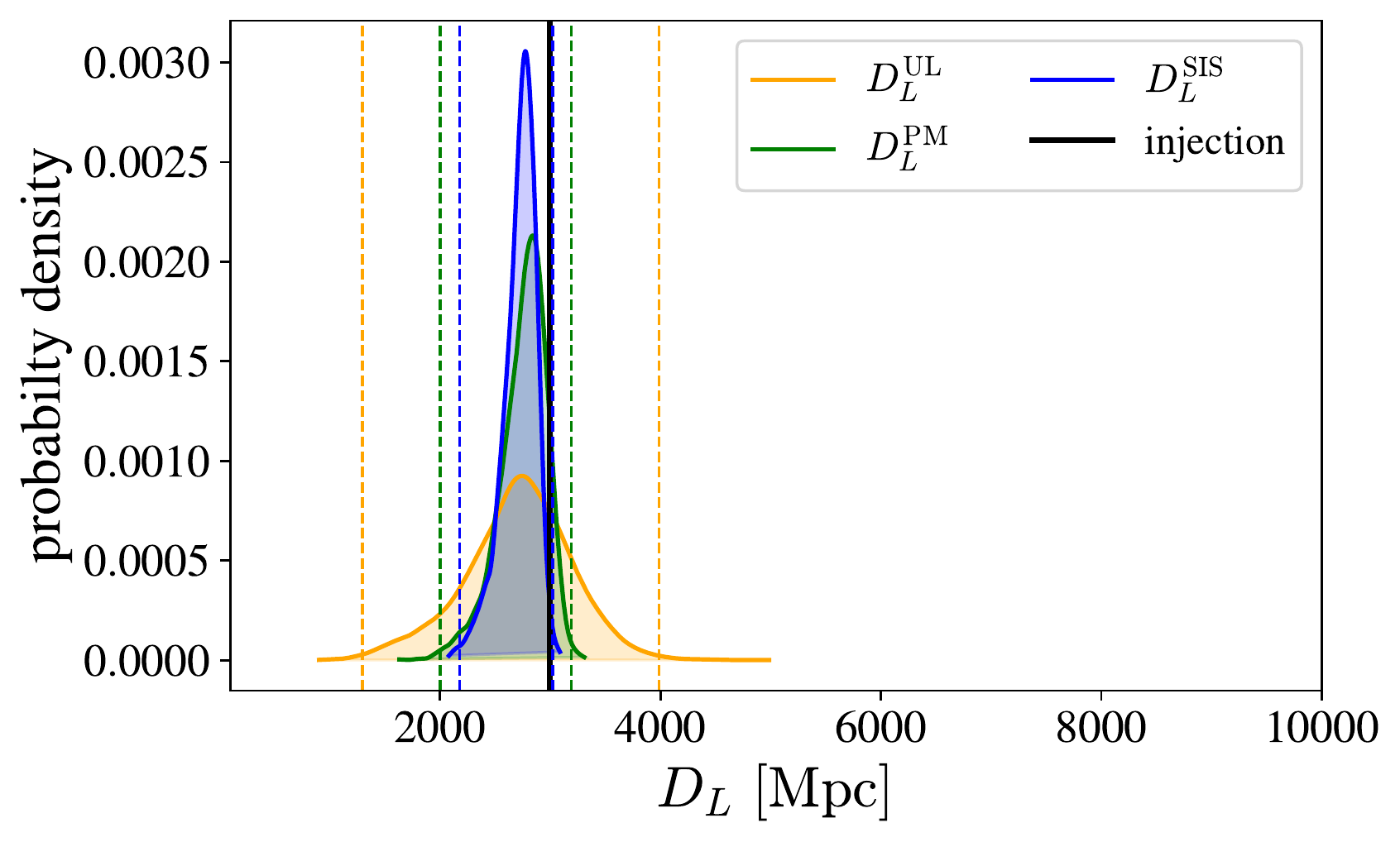}
\label{fig:reference_dl_zero_hlv}
}
\subfigure[\; HLV O3a PSDs with a Gaussian noise realization]{\includegraphics[width=.48\linewidth]{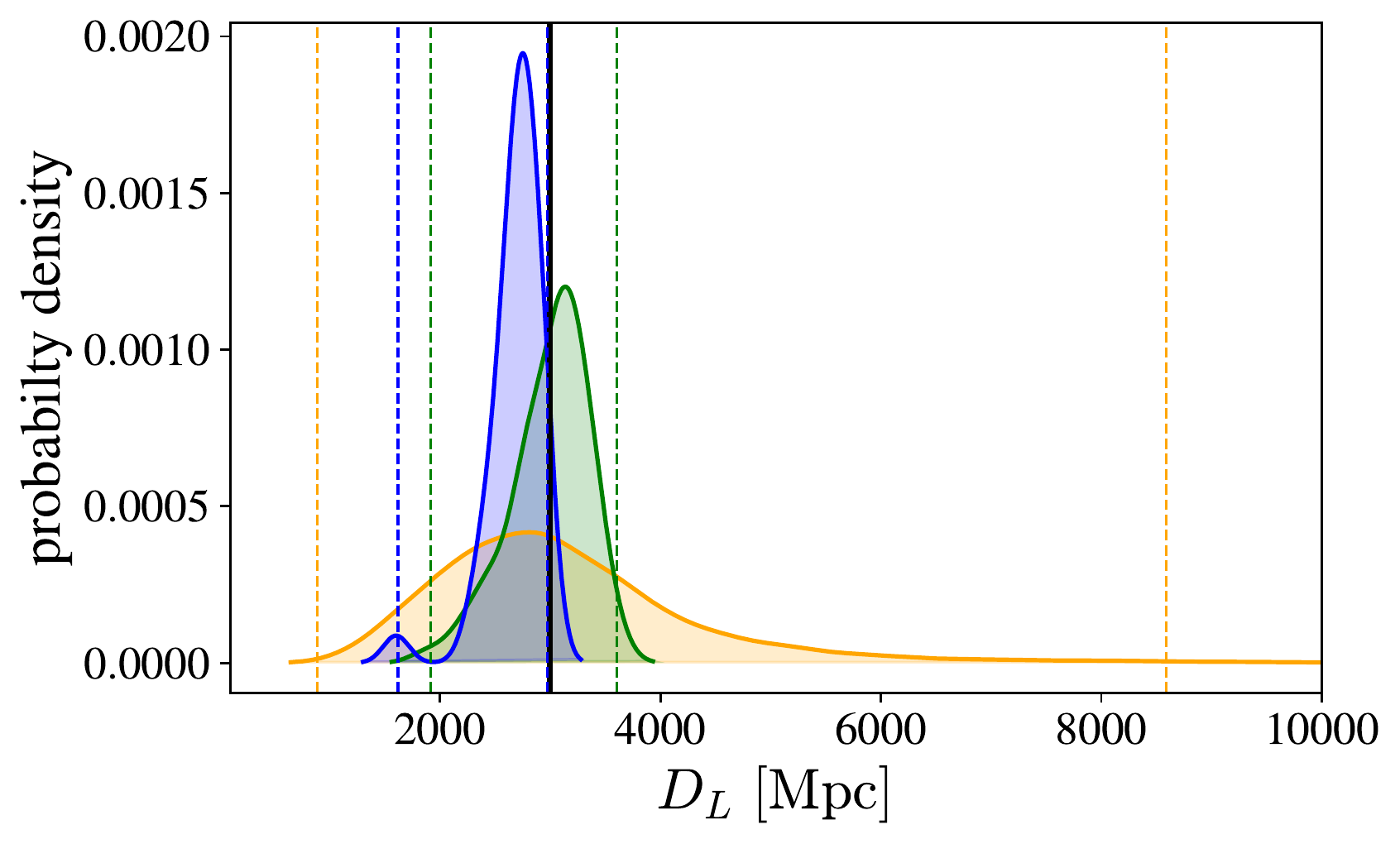}
\label{fig:reference_dl_o3a_hlv}
}
\caption{One dimensional posterior PDFs of $D_L$ recovered from $h^\mathrm{I}_{l, \textrm{PM}}(f)$ (green) and $h^\mathrm{I}_{l, \textrm{SIS}}(f)$ (blue) of the injected equal-mass BBH. For comparison, the distance posterior from an unlensed signal $p(D^\textrm{UL}_{L})$ is also shown in orange solid lines. The left panel shows the results obtained from the HLV design PSDs with zero noise. The right panel is the results from the HLV O3a PSDs involving a Gaussian noise realization. The black vertical solid lines indicate the distance to the BBH ($3~\textrm{Gpc}$). The colored vertical dashed lines indicate the lower and upper bounds of the 99\% C.I.'s for each posterior in the same color. 
\label{fig:reference_dl}}
\end{figure*}

\section{Results}
\label{sec:results}

Fig.~\ref{fig:relative} shows posteriors $p(\mu_\textrm{rel})$, $p(\mathcal{M}_+)$, and $p(\mathcal{D}_{L+})$ obtained from $h^\mathrm{I}_l$ assuming the equal-mass BBH as a source of GWs and the Case A. The posterior distributions, $p(\mu_{\rm rel})$s, obtained from both lens models accurately infer the injected values. Moreover, they are more or less identical to each other because $y=0.1$ gives $\mu_{\rm rel} \simeq 0.8$ for both models (see Fig.~\ref{fig:relation_murel_y}). The widths of posteriors, $p(\mathcal{M}_{+})$ and $p(D_{L+})$, obtained from SIS are narrower than those from PM at the 99\% C.I. In other words, the precision of PE for apparent mass and distance based on the SIS model is relatively better than that of the PM model. The differences in the ranges and widths of $p(\mathcal{M}_{+})$ can be attributed to differences in the amount of magnification ($\mu^\textrm{PM}_{+}\!\simeq\!5.5$ and $\mu^\textrm{SIS}_{+}\!=\!11$) and the corresponding SNR values ($\textrm{SNR}_\textrm{PM}\!=\!27.8$ and $\textrm{SNR}_\textrm{SIS}\!=\!39.24$). This result shows that better precision obtained from the SIS model is closely related to higher $\mu_+$ and higher SNR than those of the PM model, as expected.
Similarly, different values of $\mu_{+}$ estimated from the two models affect the distance posteriors.

Fig.~\ref{fig:reference_dl} shows the effects of strong lensing in the estimation of true $D_L$. Fig.~\ref{fig:reference_dl_zero_hlv} (Case A) shows $p(D_L)$'s obtained from the equal-mass BBH event with the HLV design PSDs and zero-noise realization. Results show that the width of $p(D_L)$ at the 99\% C.I. can be better constrained by almost a factor of up to three when both $h^\textrm{I}_l(f)$ and $h^\textrm{II}_l(f)$ are detected than that of when $h_u(f)$ is detected.
Fig.~\ref{fig:reference_dl_o3a_hlv} (Case B) shows results for the O3a PSDs with Gaussian noise, which also show improvements in the precision about a factor of up to a few (see Table \ref{tab:fig5c}).

In Table~\ref{tab:fig5c}, we compare the widths of distance posteriors at the 99\% C.I. (labeled as $\mathcal{W}_{99}$) shown in Fig.~\ref{fig:reference_dl}. We define a ratio $\mathcal{R}_{99}\!\equiv\! \mathcal{W}^\textrm{UL}_{99} / \mathcal{W}^\textrm{X}_{99}$, where $\textrm{X}$ is either unlensed (UL), PM, or SIS model. If detected, strong lensing is definitely helpful to constrain the luminosity distance better for a given detector. For example, considering the O3a PSD with Gaussian noise, $\mathcal{W}_{99}$ estimated from $p(D_L)$ is reduced by a factor of a few with respect to the results based on the unlensed signal.
In comparing the widths at the 67\% C.I., $\mathcal{W}_{67}$, we observe almost the same amount of improvements shown from $\mathcal{W}_{99}$.

\begin{table}[t]
    \caption{Quantitative comparison of the distance estimation for the equal-mass BBH obtained from unlensed, PM, and SIS models as shown in Fig.~\ref{fig:reference_dl}. The 2nd column presents $\mathcal{W}_{99}$ that is the width of $p(D_L)$ at the 99\% C.I. The 3rd column shows $\mathcal{R}_{99}$, a ratio between $\mathcal{W}_{99}$ values (see text for a definition). The 4th and 5th columns present widths and ratios at the 67\% C.I. of $p(D_L)$.} 
    \centering
    \begin{tabularx}{1.0\linewidth}{@{}X *{4}{r} @{}}
    \toprule
        \multicolumn{5}{@{}l}{\textbf{Case A: HLV design PSDs with zero-noise}}\\
        \hline 
        Signal             & $\mathcal{W}_{99}$ & $\mathcal{R}_{99}$ & $\mathcal{W}_{67}$ & $\mathcal{R}_{67}$\\
        & [Mpc] &  & [Mpc] & \\
        \hline
        Unlensed    & 2,688 & 1.00 & 923 & 1.00\\
        Lensed (PM) & 1,193 & 2.25 & 405 & 2.28\\
        Lensed (SIS)& 845 & 3.18 & 286 & 3.23\\
        \hline
        \hline\\
        \hline
        \hline
        \multicolumn{5}{@{}l}{\textbf{Case B: HLV O3a PSDs with a Gaussian noise}} \\
        \hline
        Signal & $\mathcal{W}_{99}$ & $\mathcal{R}_{99}$ & $\mathcal{W}_{67}$ & $\mathcal{R}_{67}$\\
        & [Mpc] &  & [Mpc] &  \\
        \hline
        Unlensed    & 7,700 & 1.00 & 2,074 & 1.00\\
        Lensed (PM) & 1,689 & 4.56 & 637 & 3.26\\
        Lensed (SIS)& 1,354 & 5.69 & 348 & 5.95\\
        \hline
        \hline
    \end{tabularx}
    \label{tab:fig5c}
\end{table}

In order to study the robustness of the method demonstrated in this work, we consider other mock BBHs similar to real BBHs selected from the GWTC-3~\cite{LIGOScientific:2021djp}: GW200208$\_$222617 (high mass-ratio BBH candidate), GW200216$\_$220804 (eccentric BBH candidate), and GW200219$\_$094415 (high-spin BBH candidate candidate). As shown in Table~\ref{tab:O3bBBHs}, the injection parameters for these three BBHs represent not only different populations to each other but also different from the equal-mass BBH examined thus far. However, for consistency, the same waveform model and prior distributions with the previously studied equal-mass BBH are assumed for the injected GW signals of these three additional BBHs in the PE analyses.

Table~\ref{tab:o3b-like} shows $\mathcal{W}_{99}$ and $\mathcal{R}_{99}$ values obtained for the three mock BBHs. The results are consistent for all BBHs, i.e., the distance posteriors obtained from lensed signals are better constrained than those from unlensed signals.
Differences in $\mathcal{W}_{99}$ and $\mathcal{R}_{99}$ values for the three BBHs are mainly due to the SNR values of the original unlensed signal.
For example, the network SNR of GW200208$\_$222617-like event obtained from Case A is $\textrm{SNR}_{\rm UL}^{200208} \simeq 7.4$~\cite{LIGOScientific:2021djp}. This is the lowest SNR among the four BBHs considered in this work. Its distance posterior has a larger value of $\mathcal{W}_{99}$, i.e., $\mathcal{W}_{99} \simeq 8\rm{Gpc}$ for $h_{u}$($f$) in Case A, than other BBHs. The effects of strong lensing seem to be most significant on this BBH ($\mathcal{R}^{206}_{99}\sim8$ vs $\mathcal{R}_{99}\sim3$ for other BBHs). The existence of noise is likely to suppress the effects of strong lensing in terms of SNR enhancement. Our results show that the precision of the distance estimation is still expected to be improved by strong lensing by a factor of two up to six with the existence of Gaussian noise (see Case B results in Table \ref{tab:o3b-like}).


\section{Discussions} \label{sec:discussions}
In this work, we present how to estimate the true luminosity distance $D_L$ to a BBH, supposing we can detect strongly lensed  GW signals originating from the BBH. Our results show that if an optimal strong lensing condition is provided, the posterior of $D_L$ can be better constrained by up to a factor of a few, even with the presence of simulated noise in the GW data.

For simplicity, we assume the same waveform for injection and templates in the PE analyses. Also, we consider the two simplest lens models, PM and SIS, that can produce two lensed signals in the geometrical optics limit. However, in order to consider realistic BBH populations and possible strong lensing configuration, we should consider different GW waveform models and lens models for each GW signal because the choice of the models can be attributed to the systematic biases in PE. 
Moreover, definitions and/or the relation between $y$ and $\mu_\textrm{rel}$ can be nontrivial when more lens parameters are needed to describe an amplification factor $F(f)$: For example, more than two lensed signals are expected by models such as singular-isothermal-ellipsoid~\cite{Kormann:1994} or the Navarro-Frenk-White model~\cite{Navarro:1996gj} or a more complex macrolens containing multiple microlenses~\cite{diego2019observational,PhysRevD.101.123512,mishra2021gravitational}. Therefore, future studies on more realistic and/or complicated GW lensing will be useful for understanding the effects of strong lensing in the context of GW parameter estimation in more detail.

\begin{table}[t]
    \caption{Injection parameters and SNR values of three BBHs similar to GW200208$\_$222617, GW200216$\_$220804, and GW200219$\_$094415. All parameters are median values given in Table IV of~\cite{LIGOScientific:2021djp}.} 
    \centering
    \begin{tabularx}{1.\linewidth}{@{}X *{5}{r} @{}}
    \toprule
        Event & $m_1$  & $m_2$  & $\chi_\textrm{eff}$ & $D_L$  & SNR\\
              &  [$M_\odot$] &  [$M_\odot$] &  & [Gpc] & \\
        \hline
        GW200208$\_$222617 & $51.0$ & $12.3$ & $0.45$ & $4.1$ & 7.94\\
        GW200216$\_$220804 & $51.0$ & $30.0$ & $0.10$ & $3.8$ & 10.35\\
        GW200219$\_$094415 & $37.5$ & $27.9$ & $-0.08$ & $3.4$ & 9.86\\
    \hline
    \hline
    \end{tabularx}
    \label{tab:O3bBBHs}
\end{table}

Although there has been no confirmed strongly lensed GW event from the previous observing runs~\citep{Hannuksela:2019kle,LIGOScientific:2021izm,Liu_2021,LIGOScientific:2023bwz}, searching for strongly lensed GW signals is included in the science goals of the on-going fourth observing run (O4) and future observing runs~\cite{LVK_OS_WP_22_23}. Based on forecast studies~\citep{ng2018precise,li2018gravitational,oguri2018effect,mukherjee2020probing,wierda2021beyond}, it is expected to observe $\sim\!\mathcal{O}(1)$ strongly lensed GW events per year with the design sensitivities of the aLIGO~\cite{TheLIGOScientific:2014jea} and the AdV~\cite{TheVirgo:2014hva}.
Strongly lensed GW signals are likely to be detected with better sensitivity of the LIGO-Virgo-KAGRA detector network or the third generation detectors~\cite{Piorkowska:2013eww,Cao:2014oaa} such as the Einstein Telescope~\cite{Punturo:2010zz} and the Cosmic Explorer~\cite{Reitze:2019iox}. It is worth to note that the method demonstrated in this work can be applied to any CBC sources such as a neutron star--neutron star binaries, as long as the two lensed GW signals are detected by any of the current or future detectors.

\begin{table}[t]
\caption{The $\mathcal{W}_{99}$ and $\mathcal{R}_{99}$ values that are obtained from  GW200208$\_$222617 (labeled as 208), GW200216$\_$220804 (labeled as 216), and GW200219$\_$094415 (labeled as 219)-like BBHs. Similar to Table~\ref{tab:fig5c}, results from Case A and Case B are compared for each injection. In the superscript of $\mathcal{W}_{99}$ and $\mathcal{R}_{99}$, we drop the first three and the last six digits of the event ID for convenience.
} 
\begin{tabularx}{1.\linewidth}{@{}X*{7}{c} @{}}
    \toprule
        \multicolumn{7}{@{}l}{\textbf{Case A: HLV design PSDs with zero-noise realization}}\\
        \hline 
        Signal & 
        $\mathcal{W}^{208}_{99}$ & 
        $\mathcal{R}^{208}_{99}$ & 
        $\mathcal{W}^{216}_{99}$ & 
        $\mathcal{R}^{216}_{99}$ &
        $\mathcal{W}^{219}_{99}$ & 
        $\mathcal{R}^{219}_{99}$\\
        & [Mpc] &  & [Mpc] &  & [Mpc] & \\
        \hline
        Unlensed     & 7,918 & 1.00 & 4,806 & 1.00 & 4,167 & 1.00\\
        Lensed (PM)  & 1,274 & 6.21 & 1,786 & 2.69 & 1,146 & 2.68\\
        Lensed (SIS) & 948   & 8.35 & 1282  & 3.75 & 1,147 & 3.63\\
        \hline
        \hline\\
        \hline
        \hline
        \multicolumn{7}{@{}l}{\textbf{Case B: HLV O3a PSDs with a Gaussian-noise realization}} \\
        \hline
        Signal & 
        $\mathcal{W}^{208}_{99}$ & 
        $\mathcal{R}^{208}_{99}$ & 
        $\mathcal{W}^{216}_{99}$ & 
        $\mathcal{R}^{216}_{99}$ &
        $\mathcal{W}^{219}_{99}$ & 
        $\mathcal{R}^{219}_{99}$\\
        & [Mpc] &  & [Mpc] &  & [Mpc] & \\
        \hline
        Unlensed    & 4,373 & 1.00 & 7,675 & 1.00 & 3,961 & 1.00\\
        Lensed (PM) & 1,206 & 3.63 & 1,882 & 4.08 & 2,133 & 1.86\\
        Lensed (SIS)&   637 & 6.86 & 1,693 & 4.53 & 1,437 & 2.76\\
    \hline
    \hline
\end{tabularx}
\label{tab:o3b-like}
\end{table}

More precise and accurate distance estimation is invaluable not only for understanding the formation and evolution of various CBC populations but also for constraining a Hubble constant $H_0$. When strongly lensed GW signals are detected, the PE procedure presented in this work can be used to estimate $D_L$ of CBCs with better precision. This can be helpful to better constrain the Hubble constant following the Hubble–Lema\^{i}tre law $H_0\!=\!v/d$~\cite{Lemaitre:1927, Hubble:1929}. Here, $v$ is the recessional velocity of an astronomical source and $d$ (or $D_L$) is the distance to the source that can be estimated by GW observations. Many studies have discussed methods and implications of $H_0$ measurement enabled by observing GWs from CBCs~\cite{Schutz:1986gp, Holz:2005df, Dalal:2006qt,Nissanke:2009kt,Nissanke:2013fka, Vitale:2018wlg, Borhanian:2020vyr, Borhanian:2022czq, Ghosh:2022muc}. 
GW170817~\cite{LIGOScientific:2017vwq} is the most successful example for being used to estimate $H_0$ using a GW jointly observed with its electromagnetic (EM) counterpart~\cite{LIGOScientific:2017adf}. 
Therefore, more precise distance estimation with strongly lensed GWs can shed light on the Hubble tension in the next decades.


\begin{acknowledgments}
This work is supported by the National Research Foundation of Korea (NRF) grants funded by the Ministry of Science and ICT (MSIT) of Korea Government (NRF-2020R1C1C1005863, NRF-2021R1F1A1062969, and NRF-2021M3F7A1082056). The work of K.K. is partially supported by the Korea Astronomy and Space Science Institute under the R\&D program (Project No. 2024-1-810-02) supervised by the MSIT. E.S. is partially supported by grants from the Research Grants Council of the Hong Kong (Project No.~CUHK~24304317), The Croucher Foundation of Hong Kong, the Research Committee of the Chinese University of Hong Kong, and the Science and Technology Facilities Council (Grant No.~ST/L000946/1). We are grateful for computational resources provided by the LIGO Laboratory and supported by the National Science Foundation Grants PHY-0757058 and PHY-0823459.
K.K. and E.S. equally contributed to this work as co-first authors.
\end{acknowledgments}

 \newcommand{\noop}[1]{}
\end{document}